# Phases of the Net-zero Energy Transition and Strategies to Achieve It

Jochen Markard and Daniel Rosenbloom

## Abstract

The net-zero energy transition is an extraordinary societal challenge. It requires a swift, radical and economy wide transformation. With the aim of informing research and policy, we identify general phases of this transition and the overarching strategies that may be brought to bear in tackling this challenge. Drawing from the literature on sustainability transition studies, we depict the net-zero energy transition as a non-linear, cumulative process that involves multiple, interdependent transitions in different sectors. Future emission targets can only be reached if policymaking will play a strong role in guiding these transitions. To understand the increasing complexity of the policy challenge, we distinguish four overlapping phases of development: emergence of low-carbon innovations, transition of a single sector (electricity), transitions of multiple sectors based on low-carbon electricity, and transitions in difficult-to-decarbonize sectors. We argue that each phase comes with new policy challenges on top of the already existing ones. Finally, we discuss the merits and limitations of five general strategies for decarbonization: efficiency improvement, low-carbon electrification, low-carbon fuels, negative emissions and 'untapped demand-side approaches.' While electrification has emerged as the dominant strategy, new low-carbon fuels (e.g., based on hydrogen) but also more radical changes (e.g., substitution of carbon-intensive products or lifestyle changes) merit further attention.

## 1   Introduction

The low-carbon energy transition has entered a new phase of development as more and more governments, and private firms, are making pledges to reduce their greenhouse gas emissions to net-zero. As of 2021, over 120 countries, which together represent 61% of the global greenhouse gas (GHG) emissions, had announced commitments to reach net-zero by midcentury or soon after (ECIU, 2021). Major emitters such as the United States, China, the European Union, the United Kingdom, Canada, and Japan are also on board.

With the rise of net-zero targets, societal and policy discourse surrounding climate change has shifted fundamentally. Framing the climate challenge in terms of *net-zero* foregrounds the deep changes that will be required. It will not suffice to make improvements in some sectors (e.g., phasing out coal-fired power generation in favor of renewable energy). Instead, *all GHG emissions in all sectors and places* will need to be cut or compensated for. To achieve this, far-reaching and economy wide changes in production and consumption systems will be necessary, including the transformation of 'difficult-to-decarbonize' industries such as aviation, shipping or cement production (Davis et al., 2018; Miller et al., 2021).

This analysis focuses on the energy dimensions of this challenge as energy-related $CO_2$ emissions account for about three quarters of all emissions, or 34 Gt in 2020 (IEA, 2021). Herein, the associated technological, organizational, political and institutional change processes toward eliminating or compensating for these emissions will be referred to as the *net-zero energy transition*. It includes multiple, interconnected socio-technical transitions across a broad range of sectors, from electricity to transport, buildings and industry.

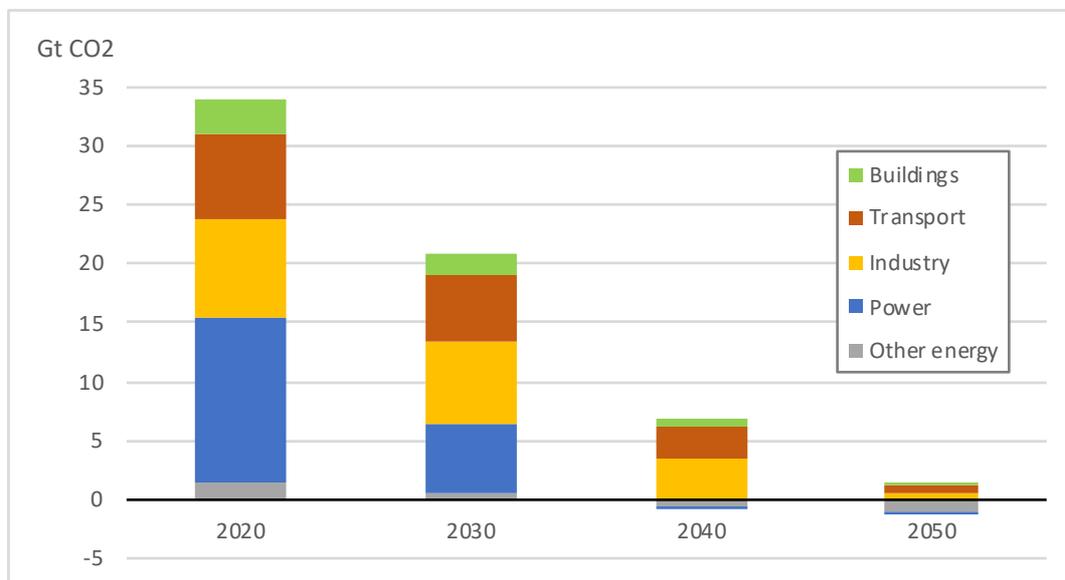

**Figure 1:** Global $CO_2$ emissions across sectors and projected path to net-zero;
Source: IEA 2021
The IEA scenario expects 'negative emissions' to occur in later years through carbon dioxide removal technologies. We provide some more detail on this in section 3.3.

Much of what we know about change processes of this sort has come from the field of transition studies (Markard et al., 2012). This research has uncovered the multiple interacting factors that shape transitions in multiple societal domains such as transport (Geels, 2005; Geels et al., 2012), and electricity (Foxon, 2013; Verbong and Geels, 2007). However, this body of work has predominantly focused on single sectors or the emergence of one major innovation. Take, for example, the shift from sailing to steam ships or from propeller to jet engines (Geels, 2002; Geels, 2006). Key frameworks in transition studies such as the multi-level perspective (Geels, 2019) or technological innovation systems (Markard, 2020) have been developed to explain transformation processes around a focal innovation or focal sector but they are not geared toward the complexity of multiple innovations and multiple transforming sectors. While scholars have identified specific deficits in existing approaches (Papachristos et al., 2013; Rosenbloom, 2020) and made first suggestions of how to overcome them (Andersen and Markard, 2020; Geels, 2018; Schot and Kanger, 2018), we still lack an overarching framework to capture transitions as complex as the net-zero energy transition. In particular, we still have a limited understanding of the new policy challenges that arise as the transition enters new phases of development, especially as change processes accelerate, accumulate, and broaden in scope.

In the following, we address some of these shortcomings. We develop a simple schematic model, which conceptualizes the net-zero energy transition as a set of interdependent sectoral transitions and distinguishes four qualitatively different phases of development. The distinction of phases is important because each comes with new, additional challenges for policymaking. Next to the phases and policy challenges, we discuss different decarbonization strategies such as low-carbon electrification, new types of fuels or lifestyle changes. These are an important element in a framework for the net-zero energy transition because the challenges that lie ahead will, most likely, require a mix of approaches to tackle decarbonization.

Our arguments build on insights from the field of sustainability transitions research, which highlights that large-scale transformations are systemic, non-linear and involve socio-political as well as techno-economic processes (Geels et al., 2017; Köhler et al., 2019; Markard et al., 2012). The phases and strategies we discuss will be of a broad, partly stylized nature to provide general guidance. Our propositions are inspired by future necessities as well as past developments in some places. Countries such as Germany, the UK or Denmark, for example, which have traditionally very much relied on electricity from fossil fuels, exhibit certain commonalities in their low-carbon transition of the electricity sector and beyond (Geels et al., 2016; Markard, 2018). We acknowledge that there is no general blueprint of the net-zero energy transition but that it very much depends on specific contextual conditions (e.g. emerging vs. industrialized economies, availability of natural resources, political priorities, industry structure, societal preferences etc.). Nonetheless, we believe that the reflections we share are valuable in situating individual transition processes in terms of an overarching pattern of development we refer to as the net-zero transition.

In section 2, we briefly introduce the perspective of sustainability transitions. Section 3 represents the core of the paper, in which we discuss the particularities of the net-zero energy transition, including four main phases (3.1), key policy challenges (3.2) and five generic decarbonization strategies (3.3). Section 4 generates lessons and offers concluding remarks.

## 2 A sustainability transition perspective

Grand sustainability challenges, from climate change to biodiversity loss and societal inequality, are pervasive and seemingly intractable crises that resist conventional policy approaches (Levin et al., 2012). They are highly complex, span across sectors and places (requiring a high degree of coordination), face conflicting views and interests of stakeholders (regarding problem definition and potential solutions) and change over time (moving target). In addition, they undermine the very basis of our (co-)existence.

Due to these particularities, grand sustainability challenges call for novel approaches in research and policy. Grounded in a growing evidentiary base, the field of sustainability transitions offers such an approach (Köhler et al., 2019). In particular, this body of research (1) demonstrates that fundamental changes in existing socio-technical systems (around electricity, transportation and buildings, for instance) are needed to address sustainability challenges; and (2) provides lessons for accelerating transition processes (Markard et al., 2020).

Transition studies take *socio-technical systems* as the primary unit undergoing change. These systems consist of different kinds of elements, including actors (e.g., firms, NGOs), institutions (e.g., policies, societal norms), technologies and infrastructures (Rip and Kemp, 1998). Below, we look at large-scale socio-technical systems[1] that provide key societal functions such as energy supply, transportation or housing (Konrad et al., 2008). In established socio-technical systems, the various elements have co-evolved over long time-spans. Dominant designs, sunk costs and vested interests reinforce a particular 'way of doing things' – a phenomenon often referred to as *lock-in* (Berkhout, 2002; Unruh, 2000). As a consequence, mature socio-technical systems are highly resistant to radical changes. In the energy sectors, we currently see how difficult it is to break up the lock-in around fossil fuels (Trencher et al., 2020).

Nonetheless, systems can and do change. Fundamental changes of socio-technical systems are referred to as *socio-technical transitions* (Kemp et al., 2001). Examples of past transitions include the transition from an oceanic shipping system based around sail to one relying on steam powered vessels (Geels, 2002) or a transport system based on horse-drawn carriages to one anchored around internal combustion automobiles and the use of oil (Geels, 2005). And, how each of these involved co-evolving changes in infrastructure (ports and roads), business models (more accurate shipping times and automotive maintenance services), practices (recreational travel), rules (speed limits), and so on.

Innovations are central to socio-technical transitions and as a response to grand challenges. Innovations include new technologies but also non-technical novelties (e.g., changes in policies, practices or lifestyles). Often, technical and non-technical changes are intertwined. This is why the term *socio-technical configurations* is often used to embrace innovations (e.g. solar-PV) but also their interconnected business models, regulations, or practices (e.g. self-consumption of electricity generated at home). Emerging socio-technical configurations such as those taking hold around renewable electricity production are expected to be the seeds for the formation of alternative and potentially more sustainable socio-technical systems (e.g., around low-carbon electricity).

Research highlights that transitions unfold in a non-linear way in the form of an S-curve (Rotmans et al., 2001; Figure 2). At an early stage, progress is slow, changes are minor and confined to niches, in which multiple innovations develop (Kemp et al., 2001; Smith and Raven, 2012). In the take-off phase, one or more innovations start to diffuse. The diffusion stimulates further improvements, which then speed the uptake of the novel innovation. During the acceleration phase, changes accumulate and eventually transform many elements of the socio-technical system (Markard et al., 2020; McMeekin et al., 2019). Finally, dynamics slow down again as a new, reconfigured system emerges and stabilizes (Geels, 2002). While this general pattern of development has mostly been associated with transitions of single socio-technical systems, we also see a similar dynamic in transitions that span across several sectors (Schot and Kanger, 2018).

---

[1] For specific systems, we also use the term sector (e.g., transport sector) because it is very common.

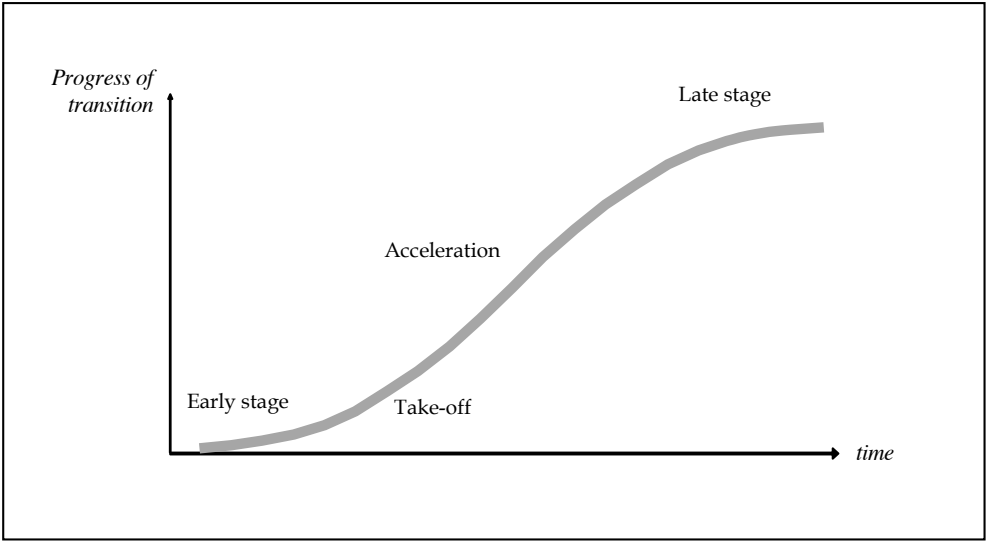

**Figure 2:** **Non-linear development and different stages of a transition**

In the context of climate change and other pressing sustainability challenges, transitions involve multiple innovations and *multiple socio-technical systems* (Andersen and Markard, 2020; Papachristos et al., 2013; Rosenbloom, 2020; Schot and Kanger, 2018). For example, the transition toward low-carbon e-mobility is not just about electric vehicles but also depends on the availability of low-carbon electricity, innovations in battery technology, or the roll-out of a charging infrastructure (Köhler et al., 2020). It involves major changes in the automobile sector, the energy sector, the battery industry and in public infrastructure. Importantly, progress in one technology (e.g., batteries) can stimulate further progress in another (e.g., electric vehicles), and vice versa. Similarly, changes in one system (e.g., transformation of electricity supply based on renewable energy sources) can lead to changes in another (e.g., utility companies investing into vehicle charging stations), and vice versa. Over time, transitions in different sectors may begin to co-evolve, potentially reinforcing each other, leading to the build-up of an even larger transition that involves multiple socio-technical systems. One part of the larger transition might be the formation of an overarching strategy, or *paradigm*, for decarbonization, similar to what Schot and Kanger (2018) have called meta-rules. One such paradigmatic feature we observe at the moment is the approach to electrify as many energy applications as possible, a strategy that is very much driven by rapidly falling costs of renewable electricity generation. We will come back to this below.

Figure 3 shows two transitions that interact, each involving a variety of technologies. At a general level, we can distinguish complementary interactions as well as competition (Markard and Hoffmann, 2016; Rosenbloom, 2019; Sandén and Hillman, 2011). Returning to the electric vehicle example, a transition to low-carbon electricity supply in the electricity sector enables the electrification and simultaneous decarbonization of transport (arguably a complementary interaction). At the same time, electric mobility drives up demand for additional low-carbon electricity, which could rationalize greater investment in the buildout of renewables (complementary) but may also result in tensions among energy end-uses (e.g., mobility and indoor temperature regulation) for limited electricity supply (competitive).

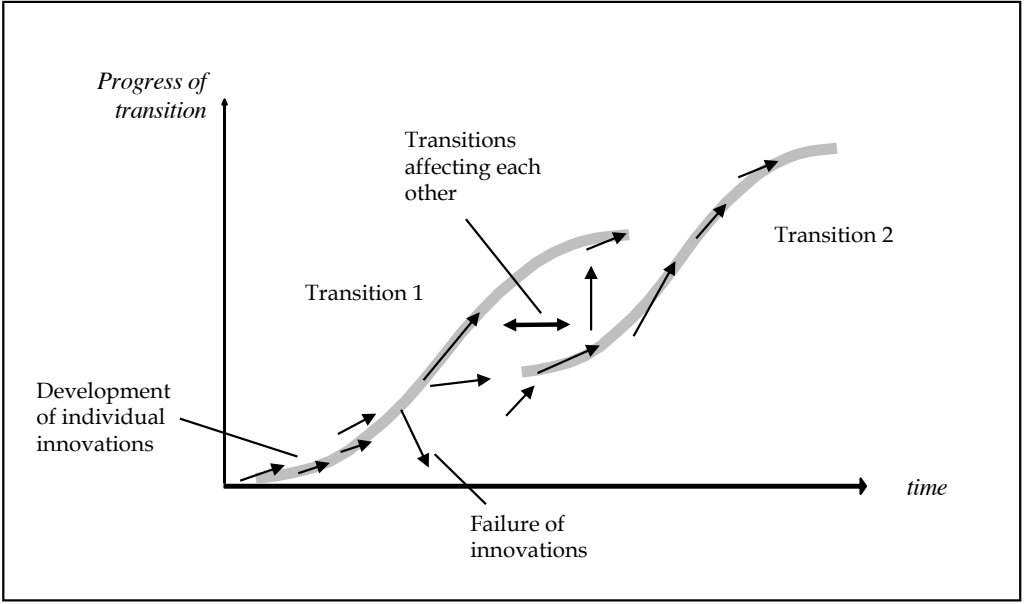

**Figure 3:** Multiple innovations and two transitions affecting each other

## 3 The net-zero energy transition

The net-zero energy transition is a particularly complex and demanding transition. Complex because it entails multiple, partly simultaneous transitions of different socio-technical systems. The transformation will affect almost every sector and part of society, there will be many interactions, unexpected developments and setbacks. Demanding because it must be swift, radical and actively pushed forward by a number of key societal actors with governments taking a central role. While past transitions often unfolded over the course of decades (Araújo, 2017; Grubler et al., 2016; Sovacool, 2016) and there was not much need for coordination between them, we only have about 30 years left to complete the net-zero transition. Different sectoral transitions therefore have to unfold in parallel, rather than one at a time. Also, it will not suffice to make incremental improvements (e.g. to just increase the fuel-efficiency of cars, or to replace coal-fired power generation by natural gas). Instead, radical changes capable of bringing about new net-zero socio-technical configurations are needed. While some innovations are readily available or in early stages of development[2], others may not even be known yet. As a consequence, the net-zero energy transition inherently involves the challenge of realizing rapid action in the context of deep uncertainty.

We understand the net-zero energy transition as an assemblage of interdependent and potentially complementary transitions that unfold in different sectors (and places) at different times and with different dynamics (Figure 4). It is a cumulative process involving multiple socio-technical transitions (gray S-curves) that together move society toward complete decarbonization. Each individual socio-technical transition entails the development and diffusion of multiple innovations (small black arrows in Figure 3, not depicted in Figure 4). These innovations may stimulate other innovations or otherwise enable transitions within adjacent socio-technical system (e.g., a disruptive business model that emerged in one sector is taken up and adapted to another sector). In other words, one transition may benefit from another (e.g., electric vehicles using low-carbon electricity), and vice versa (black arrows in Figure 4). We contend that the cumulative nature and directionality of multiple transitions is key for the ultimate realization of net-zero in a timescale needed to avert serious climate disruption (dashed gray S-curve in the background of Figure 4).

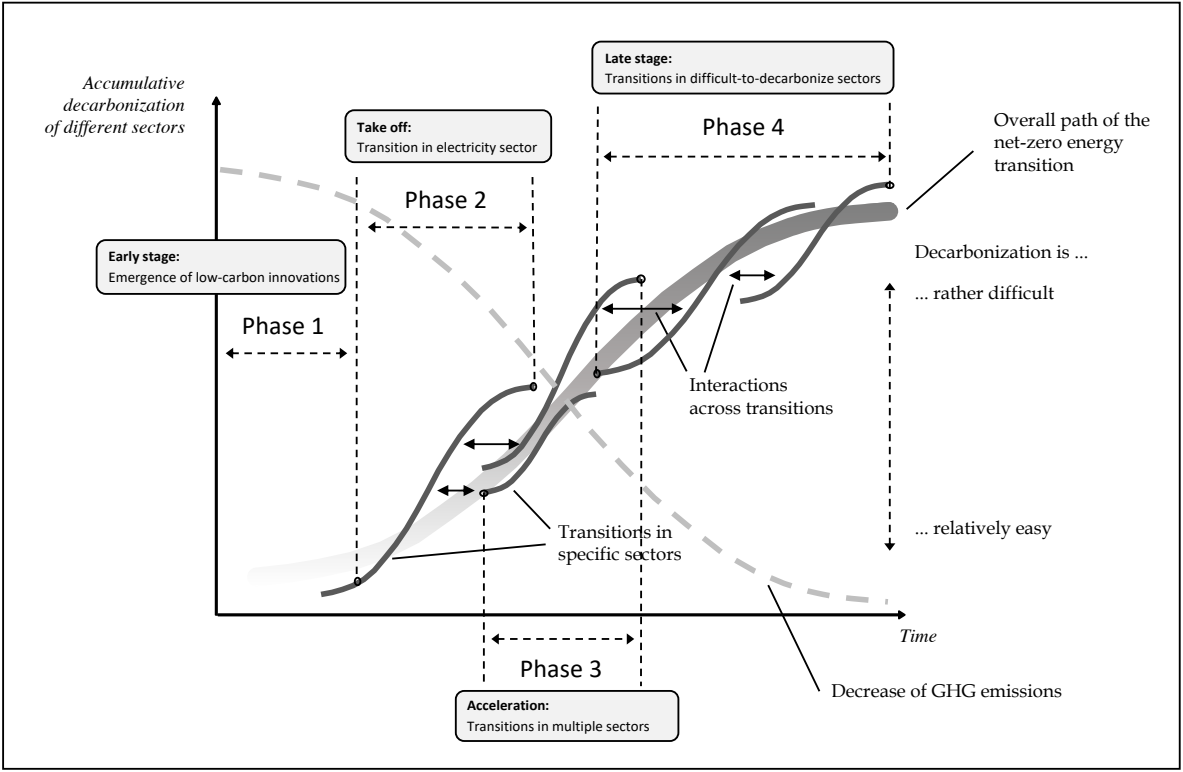

**Figure 4:** Low-carbon innovations and individual transitions build toward net-zero
Schematic model of interdependent innovation dynamics and sectoral transition. Dynamics of decline have been omitted for the sake of simplicity.

---

[2] The IEA estimates that more than 40% of the low-carbon technologies needed by 2050 are only in a developmental stage at the moment (IEA, 2021. Net Zero by 2050: A Roadmap for the Global Energy Sector. International Energy Agency, Paris, p. 224.).

This is a simple model of a larger transition that spans multiple systems and involves multiple innovations. It highlights that the overall outcome depends on how individual transitions complement each other, i.e. how their effects accumulate (e.g., the second and the third transition can use low-carbon electricity so they already start at a higher level) and how innovations, policy approaches, business models, or practices cross over from one transition to another (horizontal black arrows).[3] It also conveys the message that the transition moves from easier targets to more difficult ones, which means that policymaking will need to continually adapt and develop new solution strategies as the transition unfolds.

### 3.1 Phases to net-zero

To capture the increasing complexity of the net-zero energy transition, we distinguish four qualitatively different phases of development. This distinction is based on sectoral scope (which was wide in the first phase, then narrow with a focus on the electricity sector, then widened again) and the relative weight placed on specific decarbonization approaches. Renewable energy electrification, for instance, has emerged as such a dominant approach[4]: it is now used to decarbonize applications in the building and transport sectors and it might also be increasingly applied elsewhere. The phases are understood as mutually reinforcing and partly overlapping. To be sure, this exercise is necessarily stylized and we acknowledge that specific places or sectors may show different patterns (e.g., due to geographic, economic or political particularities).[5]

With each phase, new processes come on top of those that are already underway (Table 1). This increasing complexity has repercussions for the political dynamics that unfold and for the resulting policy challenges. Next, we discuss each phase in detail.

The *first phase* started around the 1980s and stretched into the 2000s. It was characterized by the emergence of (i) climate change as a topic of societal concern (e.g., foundation of the Intergovernmental Panel on Climate Change (IPCC); first IPCC report in 1990) and (ii) a broad range of low-carbon technologies in sectors such as electricity, buildings and transport. Examples include power generation technologies such as wind or solar-PV, new heating technologies such as solar heating or heat pumps, building efficiency technologies, biofuels, and first generation electric and fuel cell vehicles in transport. The first phase also saw changes in regulations such as grid access and feed-in regulations for independent power producers, stricter building codes, or fuel efficiency standards for vehicles. Various policies were implemented to support innovation, technology development and diffusion. New actors emerged including project developers, energy service providers or technology producers (e.g., for wind and solar). There was contestation around whether the emerging innovations would become viable alternatives and about the gravity of climate change as a policy issue. Altogether, phase 1 did not upset the equilibrium of established systems. Lock-ins remained strong and incumbents successfully weakened attempts at stringent climate policies (Meckling, 2011). Many innovations were not successful (e.g., electric vehicles) or remained confined to niches (e.g., biofuels). However, some innovations matured and started to diffuse more widely. In the electricity sector, especially wind and solar began to present promise.

---

[3] Of course, there might also be unwanted developments such as new, energy-intensive products or sectors emerging such as SUVs or space tourism (Markard, J., van Lente, H., Wells, P., Yap, X.-S., 2021. Neglected developments undermining sustainability transitions. Environmental Innovation and Societal Transitions.).

[4] The IEA estimates that by 2050, two thirds of total energy supply will be from renewable energy sources and that low-carbon electricity will be a key pillar in the pathway to net-zero: „Ever-cheaper renewable energy technologies give electricity the edge in the race to zero." (IEA 2021, p.14).

[5] Also, countries such as Norway, Switzerland or Iceland whose electricity systems were traditionally based on low-carbon electricity will have no need for a low-carbon transition in electricity (phase 2).

**Table 1: Phases of the net-zero energy transition**

|  | Phase 1<br>Early stage: Emergence of low-carbon innovations | Phase 2<br>Take-off: Transition in electricity sector | Phase 3<br>Acceleration:<br>Transitions in multiple sectors | Phase 4<br>Late stage:<br>Transitions in difficult-to-decarbonize sectors |
|---|---|---|---|---|
| **Estimated time interval** | ~ 1980 – 2010 | Since ~2010 | Since ~2015 | Since ~2020 |
| **Phase description** | Emergence of climate change issue and various low-carbon innovations | New renewable energy sources drive the transition in the electricity sector | Low-carbon electricity diffusion and complementary innovations (e.g. EVs) drive transitions in buildings and transport | Development and diffusion of new innovations (e.g. around hydrogen) to tackle applications low-carbon electricity cannot reach |
| **Sectoral scope** | Multiple sectors | Focus on the electricity sector | Multiple sectors | All sectors (also: redefinition of sectoral boundaries) |
| **Dominant decarbonization strategy** | None | Low-carbon electrification emerges as a dominant approach for decarbonization | 'Electrify everything:' Cross-sector diffusion of dominant strategy | Alternative strategies needed as limits of electrification are reached |
| **Key transition processes** | Emergence of innovations | | | |
| | | Change of entire socio-technical system, incl. decline of established configurations | | |
| | | | Accelerated diffusion and multi-system interaction | |
| | | | | Multiple, overlapping transitions and strategies for decarbonization |
| **Political dynamics** | New entrants (e.g., start-ups) and incumbent actors developed new technologies & business models<br><br>First climate policies introduced<br><br>Disagreement over the long-term viability of alternatives and the adequate reaction to climate change | Incumbent actors face increasing pressure; decline of (some) established business models (e.g., coal phase-outs)<br><br>Policies for deployment and decline in electricity proliferate<br><br>Actors struggle over renewable energy deployment, policy ambition and the pace of the transition | Organizations increasingly active across sectoral boundaries<br><br>Policies to drive electrification of buildings and transport proliferate<br><br>Opposition to renewables and climate policy continues. Increasing cross-sectoral struggles among incumbents to capture market share. | Organizations formulate individual net-zero targets and strategies]<br><br>Net-zero policy targets rapidly expand<br><br>Potential contests surround role of different energy carriers and import dependencies |
| **Policy challenges** | Advancing innovations through research, development, and demonstration | | | |
| | | Diffusing innovations and promoting the decline of carbon-intensive arrangements | | |
| | | | Policy coordination and coherence to ensure alignment across sectors, files, mandates | |
| | | | | Preventative measures, policies to monitor new energy demands |

The *second phase* began around 2010 when wind and solar-PV and other renewables started to diffuse rapidly and in many places around the world (Figure 5). Solar-PV grew from 40 GW of installed capacity in 2010 to more than 700 GW in 2020 (a growth rate of more than 30% p.a.), while onshore wind diffused from around 178 GW to 698 GW in the same period (irena.org). Momentum around these innovations helped promote a shift in focus toward the low-carbon transition of the electricity sector (Nemet, 2019). While there were also improvements in other sectors (e.g., efficiency improvements and biomass use in heating)[6], they were far outpaced by the changes in electricity. In fact, the transition in electricity toward renewables has become an initial case, perhaps a prototype even, of a sectoral transition toward net-zero (Markard, 2018). It includes major changes in power supply (including distributed generation), system balancing, sector-specific policies as well as new entrants and new business models (Geels et al., 2016). Incumbent actors have increasingly recognized that this transition threatens their established assets and business practices and have adopted political strategies to slow down the pace of transformation (Lauber and Jacobsson, 2016; Hess, 2014). Intense contests over the legitimacy and future of e.g., coal have ensued in this phase (Isoaho and Markard, 2020; Rosenbloom, 2018; Stutzer et al., 2021). A new phenomenon of phase 2 is that renewable electrification has started to emerge as a dominant strategy for decarbonization: wind, solar-PV and battery technology have seen dramatic cost and performance improvements (IRENA, 2021), which is why they might become dominant configurations for the net-zero energy transition. The second phase is still ongoing. It will end when electricity is fully decarbonized.

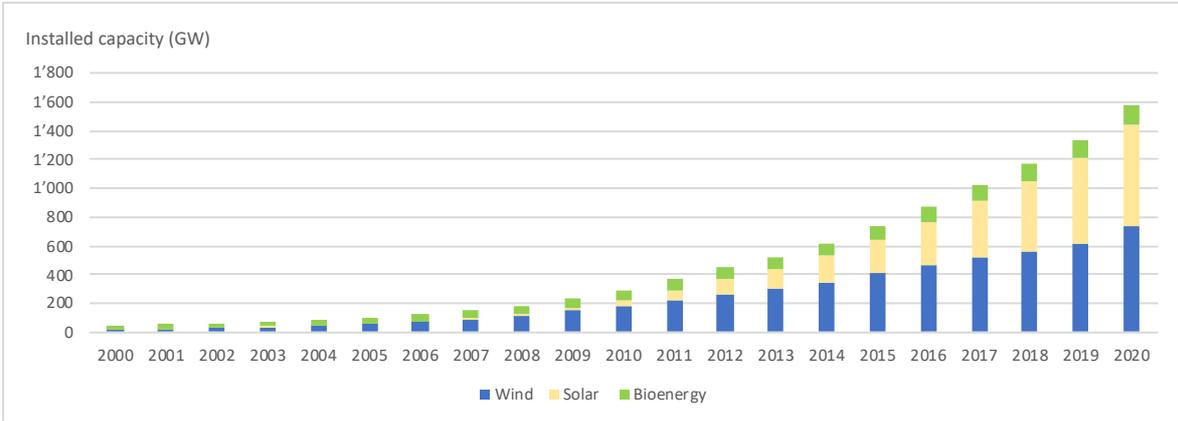

**Figure 5:** Diffusion of new renewable energy technologies (2000-2020)
Source: irena.org[7]

Phase 2 has come with several new challenges, which may also provide lessons for sustainability transitions in other sectors (Markard et al., 2020). These include whole systems change and decline. Once transitions take off and build momentum, important interdependencies among socio-technical system components are revealed (Andersen et al., 2022; McMeekin et al., 2019). This creates a need for complementary innovations (e.g., grid infrastructures, balancing technologies, demand side innovations) to ensure the functioning of the whole system (Markard and Hoffmann, 2016). Another key challenge is decline. To effectively respond to climate change, there is not only a need to expand low-carbon innovation but also to phase out existing carbon-intensive technologies such as coal-fired power generation (Rosenbloom and Rinscheid, 2020; Turnheim and Geels, 2012). Policy needs to cope with increasing resistance (Geels, 2014; Wells and Xenias, 2015) and to address e.g., regions or workers on the losing side (Johnstone and Hielscher, 2017).

---

[6] There were hardly any in transportation.
[7] https://www.irena.org/Statistics/View-Data-by-Topic/Capacity-and-Generation/Statistics-Time-Series, accessed Jan-16, 2022

The *third phase* began around 2015 and is characterized by the acceleration of electrification to decarbonize multiple sectors. The directionality of this phase is underpinned both by progress in low-carbon electricity supply (phase 2) and by complementary innovations that *use electricity* for purposes such as heating (e.g., heat pumps) or transport (e.g., batteries and electric vehicles). Renewable electrification has become a dominant paradigm and the associated decarbonization strategy is to "electrify everything" (Roberts, 2017). Also, complementary innovations are a key feature of the third phase. Phase 3 is in an early stage of development. It started to take off around 2015 with the Paris climate agreement and it will end when the potential for electrification in a broad variety of sectors has been exhausted.

In the third phase, transition dynamics are becoming more complex as they increasingly span multiple sectors or systems (Rosenbloom, 2020). These multi-sector interactions create new opportunities (e.g., cross-sectoral balancing of electricity supply and demand) but they also come with new challenges (Mäkitie et al., 2020). Norwegian utilities, for example, face a trade-off of whether to build more transmission lines to export hydroelectricity, or whether to use their hydroelectric base to drive the domestic decarbonization of transport and industry (Moe et al., 2021). The implications for incumbents are yet to become clear, though there are signs that previous incumbent strategies backing the status quo may no longer be appropriate (e.g., in light of changing consumer preferences, policy shifts, or threats from new entrants). In the building sector, for instance, there is still quite some resistance (e.g., suppliers seeking to frame natural gas as a green fuel), whereas in transport, more and more automakers are announcing ambitious targets or the end of gasoline cars (Bullard, 2021). Rising electrification will also mean the erosion of lock-ins as old assets (e.g., conventional fueling and automotive servicing stations or natural gas distribution networks) are displaced or re-envisioned. Deepening disruptions of this nature are expected to intensify political struggles where new entrants increasingly compete with incumbent service providers. Underlying these tensions are the long-term prospects for specific actor networks, business models, and social practices. And many of these debates have become increasingly polarized against a backdrop of rising right-wing populism.

The *fourth phase* started very recently when more and more jurisdictions such as the European Union introduced net-zero emission targets. The overriding emphasis during this phase is on tackling those sectors and applications that are 'difficult-to-decarbonize' (Davis et al., 2018) because of the challenges facing electrification (e.g., due to insufficient power storage capacity, remoteness of grid interconnection, or cost-efficiency considerations). Examples include aviation, shipping or long-distance trucking (Gray et al., 2021). Even though we know relatively little about phase 4 as of today, we expect that it will be characterized by a new set of low-carbon innovations. One such innovation, which is currently receiving considerable attention, is hydrogen, a general-purpose energy carrier, which can be produced in different ways and used for a broad variety of applications (Gray et al., 2021; Staffell et al., 2019). Hydrogen has the potential to play an indispensable role in driving the net-zero transition as it shows promise in replacing fossil fuels as a low-carbon feedstock in industrial processes (e.g., in chemical production and steelmaking) and as a general-purpose energy carrier (e.g., as a fuel for heavy equipment and transport). It can also interconnect with electricity to store surplus power (over long durations even) and balance intermittent renewables, potentially solidifying hydrogen and electricity as dominant energy carriers of the future (Dowling et al., 2020; van Renssen, 2020). However, hydrogen may also prolong the use of fossil fuels if natural gas is used as an input for its production. Tensions surrounding hydrogen or synthetic fuels center around production (e.g., from renewables or natural gas), use (e.g., only for difficult-to-decarbonize applications) or imports and geopolitical implications. We might also see struggles around competing alternative fuels, e.g., biofuels, hydrogen, and the reach of electrification, e.g., full battery trucks, catenary trucks, fuel cell trucks (Mäkitie et al., 2020). And while reframing the challenge in terms of net-zero along with pandemic recovery efforts appear poised to accelerate climate action, it remains too early to tell how political dynamics will unfold.

In addition to predominant supply side innovations, phase 4 might also bring about major changes on the demand side, including more serious engagement with demand reduction and changes in lifestyles (Spaargaren and Cohen, 2020). These could become increasingly necessary as the limits of supply side innovations are encountered (e.g., a reduction in overseas travel becomes a favorable solution to decrease emissions from aviation) or continued concentration on 'one side of the equation' necessitates increasingly extreme options (e.g., widescale direct air capture or bioenergy with carbon capture utilization and sequestration) with their own sustainability challenges (e.g., land use conflicts, regional environmental pollution and resource depletion). Phase 4 will end when net-zero technologies and energy carriers become dominant and targets are in sight.

Similar to phase 3, the fourth phase will be characterized by a 'multi-transition' setting. But here the interactions among previously siloed and novel energy systems are becoming even more imperative. On the one hand, electrification is continuing (and will continue) to spread rapidly across sectors and industries where fossil fuels previously dominated. And, on the other hand, alternative fuels (partly enmeshed with electricity) are now reaching nearly all parts of the economy, including difficult-to-decarbonize industries. Integrated energy planning is, therefore, essential. And even the notion of distinct energy systems (electricity, transport, hydrogen, and so on) may become obsolete. Decline will also become increasingly critical during this phase to ensure the complete displacement of fossil fuels and their associated end-use technologies. For example, communities relying on natural gas for heating will need to retire distribution networks and end-use technologies so that alternative fuel options can take their place. In contrast, blended heating systems that continue to rely on natural gas with supplemental alternative fuel injection represents a potential dead-end pathway.

*3.2 Policy challenges*

As the net-zero energy transition unfolds, policymaking becomes more and more complex. With each phase, new policy challenges come on top of the already existing ones. This evolution is comparable, and to some extent related, to how problem framings, conceptual frameworks and approaches for innovation policy changed since the 1950 from R&D policy, to systemic innovation policy and to transformative innovation policy (Schot and Steinmueller, 2018; Smith et al., 2010).

In the first phase, the focus of policymaking was on innovation. The policy challenge was to stimulate technology development and the formation of early niche markets (Hoogma et al., 2002; Kemp et al., 1998; Schot and Geels, 2008). The emphasis was on research, development, and demonstration of alternative socio-technical configurations and scholars as well as policy makers highlighted the importance of systems approaches such as technological innovation systems (Bergek et al., 2008; Markard et al., 2015). Low-carbon innovations were supported in specific sectors and places (e.g., wind in Denmark or solar in Germany) and policy programs were geared towards knowledge generation, building an early industry base and creating local markets (Bergek and Jacobsson, 2003; Dewald and Truffer, 2011; Garud and Karnøe, 2003). Even though in these early stages, there was comparatively little orientation towards transforming sectors or the entire economy, systemic innovation policy approaches will remain important in the future. The IEA states that "Reaching net zero by 2050 requires ... widespread use of technologies that are not on the market yet." (IEA, 2021; p.15) and estimates a need for public funding for R&D and demonstration of around 90 billion USD until 2030 (ibid.).

In the second phase, there was an increasing shift from innovation to deployment policies to further improve existing technologies (e.g., realizing learning effects and reaping economies of scale) and help accelerate their diffusion, e.g. through technology-specific deployment policies (Hoppmann et al., 2013; Jacobsson and Bergek, 2011; Sandén and Azar, 2005). As a complementary element in the policy mix, decline policies, particularly phase-outs (e.g., targeting coal-fired power, incandescent bulbs or internal combustion engines), are garnering increasing attention as a way to promote decarbonization (Markard and Rosenbloom, 2020; Rosenbloom and Rinscheid, 2020). These raise new issues for policy development and implementation given their implications for targeted industries and associated communities. As part of this, affected regions (around coal mining, for instance) might need to be supported, or compensated (Johnstone and Hielscher, 2017; Rinscheid et al., 2021). Overall, the policy focus is shifting towards embracing an entire sector, or whole system (McMeekin et al., 2019), thereby including a broad range of system elements (e.g., transmission grids or storage technologies in electricity) and tackling potential bottlenecks that may arise during the transition (Andersen et al., 2022; Haley, 2018).

A key policy challenge for phase 3 is cross-sectoral policy coordination (Markard et al., 2020) and multi-system interaction (Rosenbloom, 2020). This requires close exchange across governmental departments that were designed to operate individually. In earlier phases, it was possible to design decarbonization policies for one sector largely independent of those for other sectors. In the third phase, however, more and more interactions emerge across sectors, and there might also be competition between different sectors (e.g., for low-carbon electricity). Consider, for instance, how integrated energy system planning is increasingly being called for and attempted, bringing together previously siloed files on electricity, transport, buildings, and novel energy carriers (hydrogen). Phase 3 will also see an increasing need for investments into new infrastructures (e.g., public charging for electric vehicles) and public funding will play a key role in this (IEA, 2021). In fact, we might see a much stronger role of the state, and the contestation that comes along with it (Roberts and Geels, 2019), emerging in the course of phase 3.

To facilitate phase 4, continued support for innovations, sectoral transition policies and cross-sectoral policies will be needed. However, a new challenge emerges around novel technologies around hydrogen (van Renssen, 2020) or carbon capture and storage (Martin-Roberts et al., 2021) and the necessity to build up new infrastructures, e.g., to transport hydrogen or captured $CO_2$ at large scales and at a high pace (IEA, 2021). This will require massive investments, both from public and private parties (ibid.), and it will also come with a risk of potential (new) lock-ins, once these infrastructures are in place.

A very different policy challenge but also very crucial for the forth phase will be to address major changes in lifestyles and consumption patterns. This is a domain, policymaking has seen challenges in the past and, therefore, has mostly shied away from (Saujot et al., 2020). An additional policy challenge is to develop (and diffuse) a new set of low-carbon innovations (and the associated infrastructures) to tackle difficult-to-decarbonize industries and applications. Due to a rapidly growing demand for low-carbon energy carriers, we can also expect increasing political conflicts, which require careful deliberations. Conflicts may occur in energy generation (e.g., land use conflicts in the case of wind energy) and energy use (e.g., whether some applications or sectors have priority access to clean hydrogen).

## 3.3 Strategies for net-zero

There are many different approaches, or strategies, of how to decarbonize existing sectors and practices. One such strategy, which we already mentioned due to its current dominance in several sectors, centers around low-carbon electrification. In the following, we discuss net-zero strategies in some more detail. We group the large number of options into strategies and sub-strategies. Each strategy centers around a guiding principle (e.g., reducing demand, substituting fuels, changing lifestyles) and encourages the development of various socio-technical configurations that correspond to this principle (e.g., renewable power sources and electric vehicles go together with low-carbon electrification). Below, we identify five main strategies. Three strategies have already been deployed to varying degrees: energy efficiency, low-carbon electrification, and low-carbon fuels. Two additional strategies are in an early stage of development: negative emissions and a broader category with more radical demand side approaches that have remained largely unexplored so far ('untapped approaches').

The *first strategy* focuses on reducing energy demand through energy efficiency measures. This strategy has a long history and encompasses many mature innovations (e.g., LED lighting). Conservation (e.g., adjusting temperature settings for heating and cooling) and technological efficiency (e.g., building materials with higher insulation values) represent the dominant ways in which to realize this strategy. Energy efficiency measures can target all applications and sectors, though specific technologies (e.g., LED light bulbs) are often needed. The success of this strategy strongly depends on user involvement. That is, actors need to adopt more efficient technologies and change certain practices to realize gains. And of equal importance, the efficiency strategy cannot drive net-zero in isolation as there will always be some energy demand left ('residual demand').

While the efficiency strategy is relevant across all phases, it has not been given significant weight so far and largely remains underutilized.

The *second strategy* concentrates on low-carbon electrification and has emerged as a dominant approach to reach net-zero. It is composed of two sub-strategies: (2.1) substitution of carbon-intensive power generation and (2.2) extension of electrification to use cases that were traditionally served by fossil fuels (e.g., in buildings and transport). Both include a variety of socio-technical configurations, the choice of which depends on techno-economic performance improvements as well as socio-political conditions (e.g., a preference for or against nuclear for low-carbon electricity). These choices will have important implications for the operation of electricity systems. A greater reliance on distributed and intermittent technologies, for instance, may necessitate the emergence of a more flexible power system, extensive interconnections, and/or energy storage capabilities. This may also involve different ways to value energy and even flexible loads. While the second strategy has successfully been implemented, there are challenges in other sustainability dimensions such as land use or depletion of critical minerals (van den Bergh et al., 2015; Sovacool et al., 2020) and for its use in difficult-to-decarbonize sectors (Davis et al., 2018).

A *third strategy* focuses on low-carbon fuels. It reflects a growing recognition of the limitations of low-carbon electrification and seeks to substitute carbon-intensive fuels with low-carbon fuels. There are two sub-strategies. The first (3.1) is based on biofuels, which are produced from biomass such as wood, energy-crops or organic waste. Biofuels gained some traction in the 2000, showing promise in realizing GHG reductions and generating new income streams in agriculture. However, unwanted effects have also become visible – e.g., land use conflicts for food production, feedstock limitations, energy crop monocultures, or previously overlooked GHG emissions from soils or nitrous oxide (Scharlemann and Laurance, 2008). Today, biofuels remain in niche applications and have yet to play a major role in the net-zero challenge. And, given their limitations, biofuels may turn out as a dead-end pathway (Hillman and Sandén, 2008). Indeed, there is a risk that biofuels will help extend the life of internal combustion vehicles, for instance, by incrementally reducing emissions but failing to yield the radical change needed to reach net-zero.

The second sub-strategy (3.2) involves the use of hydrogen and hydrogen-based fuels. Like biofuels, hydrogen received considerable attention in the early 2000s. There were high expectations around fuel cell technology in transport (Budde et al., 2012). These expectations, however, have yet to be borne out and interest largely faded until recently. Net-zero targets have created renewed interest in hydrogen as an energy carrier that might be used to address many difficult-to-decarbonize sectors (e.g., shipping, aviation, and others). Currently, many hydrogen-based technologies are in an early stage of development (except hydrogen vehicles) and it heavily depends on political support. Major limitations include energy losses in production and conversion, and high costs. There are also risks that hydrogen – when produced from natural gas – locks in continued reliance on fossil fuels. The contribution of hydrogen to net-zero will, therefore, hinge on the uncertain prospects of carbon capture utilization and sequestration or a widespread expansion of low-carbon electricity to produce hydrogen through electrolysis.

**Table 2: Five main strategies for decarbonization toward net-zero[8]**

|  | **Principle(s)** | **Examples** | **Maturity** | **Particularities** | **Limitations** |
|---|---|---|---|---|---|
| **Energy efficiency (1)** | Conservation (1.1) | Switch off unused loads | Many mature technologies and services | Hinges on energy users | Decarbonization gap: residual energy demand cannot be addressed by efficiency |
|  | Technological efficiency (1.2) | LED lightbulbs; higher fuel economy gas engines |  | Variety of approaches & technologies tailored to different use cases |  |
| **Low-carbon electrification (2)** | Substitute carbon-intensive with low-carbon electricity generation (2.1) | Renewable power generation; nuclear; storage; flexibility technologies; | Mostly mature (but still much potential for further diffusion) | Variable renewables require more flexible grid | Land use; minerals and resource needs; many sectors (e.g. heavy transport, shipping, air travel) currently defy electrification |
|  | Electrify additional use cases (2.2) | Electric vehicles; air-source heat pumps | Diffusing rapidly in transport, more slowly in buildings | Significant expansion of low-carbon electricity needed |  |
| **Low-carbon fuels (3)** | Alternative fuels based on biomass (3.1) | Ethanol and biodiesel, methane and biogas, biomass power | Mature | Potential dead-end | Secondary GHG emissions, land use, monocultures, soil degradation, limited feedstock, etc. |
|  | Direct hydrogen use or synthetic fuels based on hydrogen (3.2) | Hydrogen, ammonia | Very early stage | Produced through a variety of approaches, including fossil fuels. Geopolitical implications | Conversion losses, high costs, potential fossil fuel lock-ins |
| **Negative emissions (4)** | CCS based technologies (4.1) | bioenergy and CCS, direct air capture and CCS | Early stage / some experience | Societal acceptance of CCS might be a barrier | Energy intensive, requires renewable electricity |
|  | Other negative emissions technologies (4.2) | Afforestation, enhanced weathering, soil carbon sequestration, restoration of peatlands | Early stage | Many nature-based strategies | Processes not (yet) fully understood; not commercially viable; politically challenging |

---

[8] Note that we don't include carbon dioxide removal (CDM) approaches here as they are not strictly related to the energy transition. Bioenergy plus carbon capture and storage (BECCS) is another technological option (perhaps to be subsumed under 3.1) we don't address here in any further detail.

|  | **Principle(s)** | **Examples** | **Maturity** | **Particularities** | **Limitations** |
|---|---|---|---|---|---|
| **Untapped demand side approaches (5)** | Major changes in lifestyles and work practices (5.1) | Car-free lifestyle; restrictions to regional air travel; telework & conferencing | Very early stage to emergence (e.g., some car-free communities) | Strong user & industry resistance, low political feasibility | Politically and administratively challenging; many other institutional changes required (e.g., standards, building codes) |
|  | Radical substitution of carbon-intensive products (5.2) | Replace cement or steel with plant-based structural materials | Very early stage | Requires radical changes in business models | Hinges on viable alternatives that generate sufficient interest and resources |
|  | Restrict emergence of new carbon-intensive practices (5.3) | SUVs; space tourism; outdoor heating | Very early stage | Requires societal debate about needs and values | Institutional capacity building needed to shift from 'firefighting' to 'fireproofing' |

A *fourth strategy* centers around carbon dioxide removal and 'negative emission technologies' (Haszeldine et al., 2018). It is based on the idea that $CO_2$ needs to actively be removed from the atmosphere because other strategies are not sufficient to reach net-zero goals in time. There are two sub-strategies: carbon capture and storage (CCS) based approaches (4.1) and other negative emission strategies (4.2).

CCS technology is already used to capture and store $CO_2$ from fossil fuels before it is released into the atmosphere through various methods (pre-combustion, post-combustion, oxyfuel) and can be paired with other technologies to achieve negative emissions (Martin-Roberts et al., 2021). For example, when pairing bioenergy with CCS (BECCS), biomass can act as a carbon sink while it grows and when converted to bioenergy, CCS comes in handy. A second approach can be through direct air capture (DAC) and CCS. DAC technology is one of few that can remove $CO_2$ directly from the atmosphere. CCS has been under development for more than 25 years and, so far, it has fallen short compared both to earlier expectations (e.g., in terms of costs and performance) and future necessities (e.g., in terms of required capacities) (Martin-Roberts et al., 2021).

Other negative emission strategies include reforestation (planting trees where forests used to be) and afforestation (planting trees where there were previously none), methods like enhanced weathering or ocean alkalinity enhancement (spreading fine basalt rock over large land/sea areas to accelerate chemical weathering reactions leading to $CO_2$ removal and storage in e.g., solid carbonate minerals, Bach et al., 2019), and carbon sequestration by changing agricultural practices (away from conventional farming towards e.g., agroforestry). These strategies are in an early stage of development and many of the underlying processes and implications are not yet fully understood.

While strategies 1-4 may contribute significantly to the pathway to net-zero, it is not yet possible to discern all approaches needed for full decarbonization by mid or end of century. Acknowledging this, we offer three illustrative examples of relatively untapped approaches that can complement more established solutions (e.g., should limitations be encountered around electrification or low-carbon fuels). These include: major lifestyle changes (5.1), radical substitution of products (5.2) and restricting new carbon-intensive practices (5.3).

While strategy 1 already targets demand side changes, much more profound changes in lifestyles may be required to reduce energy demand to levels compatible with net-zero emissions. Consider, for instance, car-free housing projects (supported by new approaches in city planning), telework and virtual conferencing to reduce travel, low-carbon diets, local vacations, or extensive changes in consumption patterns (e.g., a shift away from fast fashion). A second approach (5.2) centers around the substitution of carbon-intensive, difficult-to-decarbonize products such as cement, steel or aluminum with alternative materials (e.g., wood or other plant-based building materials). A third approach (5.3) focuses on preventing or downscaling unwanted developments. The ongoing diffusion of pickup trucks and SUVs, for example, significantly increases energy demand in transportation and sets even higher hurdles for electrifying transport. There are also entirely new energy uses emerging. Think of space tourism: if current pilot projects scale up and – eventually – diffuse more widely (as air travel did some decades ago), they will make achieving net-zero emission targets all the more difficult.

So far, these strategies have remained largely unexplored or restricted to very small niches. However, it may become increasingly vital to mobilize a broader range of approaches and to gain experience with policies to support radical change. Efforts to advance these strategies can expect to encounter serious political challenges such as when the German Green Party suggested a "veggie day" in staff canteens or a restriction of suburban single-family homes (The Guardian, 2021). In Switzerland, a policy initiative to ban SUVs and off-road vehicles was rejected by parliament and government some years ago but stricter emission regulations were implemented instead (Swiss Confederation, 2008). While political feasibility may still be a major hurdle for these more far-reaching approaches to decarbonization, pressure to widen the repertoire of decarbonization approaches is likely to increase.

# 4  Conclusions

Reducing GHG emissions to net-zero is one of the biggest challenges of our times. Building on research in the field of sustainability transitions, we have proposed four major phases of change. After an early period with the emergence of a first generation of low-carbon technologies (phase 1), we are currently witnessing the acceleration of the low-carbon transition in the electricity sector (phase 2). This will provide the basis for electrifying further energy uses and sectors such as buildings and automobility, which have thus far relied on fossil fuels (phase 3). While low-carbon electrification has become the dominant strategy for decarbonization, it is also clear that this will not suffice. There are difficult-to-decarbonize sectors (aviation, shipping, cement), for which it is not a feasible option.

Engagement with climate change over the past few decades suggests that complementary strategies are beginning to move us toward net-zero, including energy efficiency, low-carbon electrification, and low-carbon fuels. However, there is also room to consider more radical strategies, from major changes in lifestyles to the substitution of difficult-to-decarbonize materials (e.g. steel, cement) and a restriction of new, carbon-intensive practices (e.g., space tourism).

To be sure, the phases and strategies we suggested are not meant to capture all possible pathways. Indeed, different developments are possible such as rapid advancements in negative emission technologies (engineered or natural) to either draw down historical emissions or offset ongoing emissions. Net-zero pathways can also be expected to vary considerably across contexts (e.g., a jurisdiction with an already near-decarbonized electric power system versus another with mixed or carbon-intensive electricity). And, there may be unwanted developments that could lead society along dead-end pathways (with unsuccessful strategies) or dystopian futures in which society fails to reach net-zero emissions.

Taken together, our analysis points to several principles for sustainability transition policy aimed at the pursuit of net-zero. First, it is important to recognize climate change as a grand challenge and systems problem, which requires fundamental transformations instead of piecemeal adaptation (Rosenbloom et al., 2020). For this, public policies have to foster radical innovations, which include new socio-technical configurations (e.g. around hydrogen) as well as non-technical innovations (e.g. low-carbon lifestyles). Second, given the inherent uncertainty in the transition toward net-zero, it will be vital to carefully monitor ongoing developments, in particular to avoid dead-end pathways (e.g. biofuels, natural gas use), which create further delays and sunk investments (Hillman and Sandén, 2008; Meadowcroft et al., 2019). Third, policymaking has to manage conflicting interests (e.g. by compensating losers), forge strong coalitions in favor of change (Hess, 2019; Meckling et al., 2015), and carefully attend to those political strategies that seek to undermine stringent decarbonization (e.g., the new politics of delay, Lamb et al., 2020). Fourth, it will be necessary to develop and apply a mix of policies, which reflect the particularities of different phases of development as well as sector and country specific conditions (Meckling et al., 2017; Rosenbloom et al., 2020). It will not be possible to devise one-size-fits-all approaches. Finally, policymaking has to prepare to tackle the inconvenient policy areas, including established but unsustainable consumption practices and lifestyles (Spaargaren and Cohen, 2020). These will evoke much opposition and resistance but they may prove to play a key role in our endeavor to successfully navigate the net-zero energy transition.

## Acknowledgements

We thank Allan Dahl Anderson, Leticia Müller, Dianne Hondeborg and our colleagues in the Pathways project and at the ETH-group for Sustainability and Technology for their comments and inputs to earlier versions of this manuscript. Jochen Markard acknowledges funding from the Norwegian Research Council (Conflicting Transition Pathways for Deep Decarbonization, Grant number 295062/E20) and from the Swiss Federal Office of Energy (SWEET program, PATHFNDR consortium). Daniel Rosenbloom would like to acknowledge the financial support of the Social Sciences and Humanities Research Council of Canada.